\documentstyle[aps]{revtex}
\draft
\begin{document}
\title{Overlap functions in correlation methods and quasifree nucleon
knockout from $^{16}$O}
\author{M.K. Gaidarov, K.A. Pavlova, A.N. Antonov, M.V.Stoitsov,
and S.S. Dimitrova}
\address{Institute of Nuclear Research and Nuclear Energy,
Bulgarian Academy of Sciences, Sofia 1784, Bulgaria}
\author{C. Giusti}
\address{Dipartimento di Fisica Nucleare e Teorica,
Universit\`a di Pavia,\\
Istituto Nazionale di Fisica Nucleare, Sezione di Pavia,
Pavia, Italy}
\maketitle

\begin{abstract}
The cross sections of the ($e,e'N$) and ($\gamma,p$) reactions on $^{16}$O are
calculated, for the transitions to the $1/2^{-}$ ground state and the first
$3/2^{-}$ excited state of the residual nucleus, using single--particle overlap
functions obtained on the basis of one--body density matrices within different
correlation methods. The electron--induced one--nucleon knockout reaction is
treated within a nonrelativistic DWIA framework. The theoretical treatment of
the ($\gamma,p$) reaction includes both contributions of the direct knockout
mechanism and of meson--exchange currents. The results are sensitive to details
of the different overlap functions. The consistent analysis of the reaction
cross sections and the comparison with the experimental data make it possible
to study the nucleon--nucleon correlation effects.
\end {abstract}
\vspace{1cm}

\section{Introduction}

Quasifree ($e,e'p$) knockout reactions have proved to be a powerful tool for
nuclear structure investigations. The large amount of data from these processes now
available \cite{FM84,Oxford,Lap93,Leuschner,Kra90,Steen88,Her88} gives detailed
information on single--particle (s.p.) aspects of nuclear structure,
revealing the properties of the nucleon--hole states contained in
the hole spectral function. In this way, it is possible to point
out the validity and limits of the s.p. description of nuclei.

The knowledge of the hole spectral function provides information
on the spectroscopic factors for the removal process and on the
momentum distribution of transitions to discrete final states of
the residual nucleus. The spectroscopic factors obtained from the
data exhibit a remarkable fragmentation over
final states, thus indicating that nucleon-nucleon (NN)
correlations are not negligible \cite{Lap93,Ant88,Ant93}. The
nucleon momentum distributions extracted for a variety of nuclei show
unambiguosly the existence of high--momentum components \cite{Cio91} which is not
the case in the mean--field approximation (MFA). They are caused by the
short--range and tensor NN correlations in nuclei which originate
from specific peculiarities of the nucleon--nucleon forces at small
distances.

Along with the experimental studies, a precise theoretical
treatment is also needed, taking correctly into account all main
ingredients of the cross sections and regulating various
approximations. Only in this way a proper comparison with
experimental data is possible and reliable nuclear structure
information can be extracted.  The experimental momentum distributions are
reproduced, with a good degree of accuracy, in a wide range of nuclei and in
different kinematics, by a nonrelativistic treatment based on the distorted wave
impulse approximation (DWIA), where also spin dependence in the final--state
interactions (FSI) and Coulomb distortion of electron waves are taken into
account \cite{Oxford,DWEEPY}. Similar approaches based on a fully relativistic
DWIA treatment are also available \cite{relat}.

The exclusive nature of the quasifree knockout reactions in which the emitted
nucleon is experimentally detected, is reflected theoretically in a transition
amplitude which can be expressed in terms of the s.p. function representing the
overlap between the target and residual nucleus states. In standard DWIA
calculations phenomenological s.p. bound state wave functions are usually
adopted, which do not include correlations. The normalization of the wave
function, that is identified with the spectroscopic factor, is fitted to the
data and its deviation from the predictions of the MFA is
usually interpreted as evidence for the presence of correlations.

Explicit calculations of the hole spectral function and of fully correlated
overlap functions for complex nuclei are very difficult. Only very recently
the first successful parameter--free comparison of experiment and theory
including the absolute normalization in $p$-shell nuclei has been performed for
the $^7$Li($e,e'p$) reaction \cite{Li7}. For heavier nuclei, the effects of a
spectral function containing short--range and tensor correlations on the
$^{16}$O($e,e'p$) reaction have been investigated in ref. \cite{PR}. The
effects of a spectral function containing long--range correlations on the same
reaction have been investigated in ref. \cite{Amir}. In both cases a fair
agreement with the shape of the experimental momentum distributions is
obtained, but the size of the experimental cross section is overestimated and
the spectroscopic factors determined by a fit to the data are thus lower than
those predicted by the calculation of the spectral function. A calculation able
to account for effects of both short--range (SRC) and long--range correlations
(LRC) is extremely difficult, since it requires excessively large model space.
A method to deal with SRC and LRC consistently has been proposed and applied
to calculate the spectroscopic factors for one--nucleon knockout from
$^{16}$O \cite{Geurts}.

It has been shown recently that absolute spectroscopic factors and overlap
functions for one--nucleon removal reactions can be extracted from the one-body
density matrix (OBDM) of the target nucleus \cite{Vn93}. The advantage of this
procedure is that it avoids the complicated task of calculating the total
nuclear spectral function. The procedure for extracting bound--state overlap
functions has been applied~\cite{Sto96,Vn97,Di97,Gai98,Gai99a} to OBDM emerging from
various correlation methods such as the Jastrow correlation method (JCM) \cite{Sto93},
the Correlated Basis Function (CBF) theory \cite{Vn97,Sa96} and the Green function
method (GFM) \cite{Po96}. It has been shown that these functions are of particular
importance, since they contain nucleon correlations which are accounted for to
different extent in the various theoretical methods considered. On the other
hand, their reliability for analyzing quantities which are sensitive to the NN
correlations can be proved. The applicability of the theoretically
calculated overlap functions has been tested in the description of
the $^{16}$O($p,d$) pickup reaction~\cite{Di97,Gai98,Gai99a} and of the
$^{40}$Ca($p,d$) reaction (within the JCM) \cite{Di97}. It has
been found a good overall agreement between the calculated and the
empirical cross sections. It has been pointed out also that
acceptable spectroscopic factors can be obtained with the method
proposed. Considering the role of the short--range and tensor correlations, it
has been concluded that the LRC corresponding to collective degrees of freedom
have to be taken also into account in order to achieve a better
agreement with the ($p,d$) data. The LRC can have sizable effects
on the spectroscopic factors, on the shape of the overlap function
and, as a consequence, on the cross sections.

In the present work the s.p. overlap functions obtained in different
correlation methods mentioned above are used firstly to calculate the cross sections
of the $^{16}$O($e,e'n$) and $^{16}$O($e,e'p$) knockout reactions. The aim of
this investigation is to clarify the importance of the effects of various types
of correlations on the reaction cross sections also in comparison with
empirical ($e,e'p$) data. The calculation of the cross sections is based on the same
nonrelativistic DWIA treatment \cite{DWEEPY} already used for the analysis of
many experimental data. Secondly, the s.p. overlap functions are also used to calculate
the cross section of the $^{16}$O($\gamma,p$) reaction. For the photon-induced
reaction we have adopted the theoretical treatment of ref.~\cite{Benenti},
where the contributions of the direct knockout mechanism (DKO) and of meson-exchange
currents (MEC) are evaluated consistently. The comparison of calculations, with
consistent theoretical ingredients and constrained parameters, for the
($e,e'p$) and ($\gamma,p$) cross sections can enable us to check, in comparison
with data, the consistency of the theoretical description of the two reactions.
Moreover, it allows us to investigate the behaviour of overlap functions from
different correlation methods in a wide range of momenta, in particular
at the large values that can be sampled in the ($\gamma,p$) reaction and where
correlation effects are expected to be more sizable.

The theoretical framework for the calculation of the ($e,e'N$) and
($\gamma,p$) cross sections is presented in Section~2, together with the
procedure to extract the s.p. overlap functions from the OBDM of the target
nucleus. A short description of the correlation methods used is also given.
The results of the calculations are presented and discussed in Section~3. The
concluding remarks are given in Section~4.

\vspace{5mm}

\section{Theoretical approach}

\subsection{The DWIA formalism for the ($e,e'N$) reaction}

The cross sections of the $^{16}$O($e,e'n$) and $^{16}$O($e,e'p$) reactions
have been calculated with the code DWEEPY~\cite{DWEEPY}, which was able to give
a good description of the ($e,e'p$) experimental momentum distributions, for
transitions to different final states, in a wide range of nuclei and in
different kinematics (see, e.g., ref.~\cite{Oxford} and, more specifically for
the analysis of the $^{16}$O($e,e'p$) reaction, ref.~\cite{Leuschner}). The
calculation is based on a nonrelativistic DWIA description of the quasifree
nucleon knockout process and accounts for both FSI and Coulomb distortion of
the electron waves. The essential features of the formalism are given in this
section. More details can be found in refs.~\cite{Oxford,DWEEPY}.

In the one-photon exchange approximation, the general expression of the
coincidence unpolarized cross section for the reaction induced by an electron,
with momentum ${\mbox{\boldmath $p$}}_0$ and energy $E_0$, with
$E_0=|{\mbox{\boldmath $p$}}_0|=p_0$, where a nucleon, with momentum
${\mbox{\boldmath $p$}}'$ end energy $E'$, is ejected from a nucleus, can be
written in terms of four structure functions $W_{\mathrm i}$, as~\cite{Oxford}

\begin{eqnarray}
\frac{{\mathrm d}^{3}\sigma}{{\mathrm d}E'_{0}{\mathrm d}\Omega'_{0}
{\mathrm d}\Omega'} = \frac{\pi e^2}{2 q}
\Gamma_{\mathrm V} \,\Omega_{\mathrm f} f_{\mathrm{rec}}
&  & [\epsilon_{\mathrm L}W_{\mathrm L}+W_{\mathrm T}+\epsilon\,W_{\mathrm{TT}}
\cos2\phi  \nonumber \\
& + & \sqrt{\epsilon_{\mathrm L}(1+
\epsilon)}W_{\mathrm{TL}}\cos\phi],
\label{eq:cs}
\end{eqnarray}
where $e^2/4\pi \simeq 1/137$, $E'_0$ is the energy of the scattered electron,
with momentum ${\mbox{\boldmath $p'$}}_0$,
and $\phi$ is the angle between the plane of the
electrons and the plane containing the momentum transfer
${\mbox{\boldmath $q$}}$ and ${\mbox{\boldmath $p$}}'$.
The quantity
\begin{equation}
\epsilon = \left(1-\frac{2{\mbox{\boldmath $q$}}^{2}}{q^{2}_{\mu}}\tan^{2}
\frac{\theta}{2}\right)^{-1}
\end{equation}
measures the polarization of the virtual photon exchanged by the electron
scattered at an angle $\theta$ and
\begin{equation}
\epsilon_{\mathrm L} = -\frac{q^{2}_{\mu}}{{\mbox{\boldmath $q$}}^{2}}\epsilon,
\end{equation}
where $q^{2}_{\mu}=\omega^{2}-{\mbox{\boldmath $q$}}^{2}$, with
$\omega=E_0-E'_0$ and
${\mbox{\boldmath $q$}}={\mbox{\boldmath $p$}}_0-{\mbox{\boldmath $p$}}'_0$,
is the four-momentum transfer. The factor

\begin{equation}
\Gamma_{\mathrm V} = \frac{e^2}{8\pi^3} \, \frac{E'_0}{E_0} \,
\frac{q}{q^{2}_{\mu}} \, \frac{1}{\epsilon-1},
\end{equation}
is the flux of virtual photons, $\Omega_{\mathrm f} =p'E'$ is the phase-space
factor and
\begin{equation}
f_{\mathrm rec}^{-1} = 1- \frac{E'}{E_{\mathrm B}}
\frac{{\mbox{\boldmath $p$}}'\cdot {\mbox{\boldmath $p$}}_{\mathrm B}}
{{{\mbox{\boldmath $p$}}'}^{2}}
\end{equation}
is the inverse of the recoil factor. The quantity $E_{\mathrm B}$ is the
total relativistic energy of the residual nucleus with momentum
${\mbox{\boldmath $p$}}_{\mathrm B}={\mbox{\boldmath $q$}}-
{\mbox{\boldmath $p$}}'$.

The structure functions represent the response of the nucleus to the
longitudinal ($\mathrm L$) and transverse ($\mathrm T$) components of the
electromagnetic interaction. They are obtained from suitable combinations of
the components of the hadron tensor~\cite{Oxford} and are thus given by
bilinear combinations of the Fourier transforms of the transition matrix
elements of the nuclear charge-current density operator taken between initial
and final nuclear states
\begin{equation}
J^{\mu}({\mbox{\boldmath $q$}}) = \int \langle\Psi_{\mathrm{F}}
|\hat{J}^{\mu}({\mbox{\boldmath $r$}})|\Psi_{\mathrm{I}}\rangle
{\mathrm{e}}^{\,{\mathrm{i}}
{\footnotesize {\mbox{\boldmath $q$}}}
\cdot {\footnotesize {\mbox{\boldmath $r$}}}}
{\mathrm d}{\mbox{\boldmath $r$}} .
\label{eq:jm}
\end{equation}
These integrals represent the basic ingredients of the calculation.

The DWIA treatment of the matrix elements in Eq.~(\ref{eq:jm}) is based on the
following assumptions :

i) An exclusive process is considered, where the residual nucleus is left in a
discrete eigenstate $|\Psi^{\mathrm B}_\alpha(E) \rangle$ of its Hamiltonian,
with energy $E$ and quantum numbers $\alpha$.

ii) The final nuclear state is projected onto the channel subspace spanned by
the vectors corresponding to a nucleon, at the position
${\mbox{\boldmath $r$}}_1$, and the residual nucleus in the state
$|\Psi^{\mathrm B}_\alpha(E) \rangle$. This assumption is justified by the
asymptotic configuration considered for the final state.

iii) The nuclear-current operator does not connect different channel subspaces.
Thus, also the initial state is projected onto the selected channel subspace.
This assumption is the basis of the direct knockout mechanism.

Then, the transition matrix elements in Eq.~(\ref{eq:jm}) can be written in a
one-body representation as
\begin{equation}
J^{\mu}({\mbox{\boldmath $q$}}) = \int \chi^{(-)*}_{E\alpha}
({\mbox{\boldmath $r$}}_1)
\hat{J}^{\mu}_{\mathrm{eff}}({\mbox{\boldmath $r$}},
{\mbox{\boldmath $r$}}_1) \phi_{E\alpha}({\mbox{\boldmath $r$}}_1)
\left[S_\alpha(E)\right]^{1/2}
{\mathrm{e}}^{\,{\mathrm{i}} {\footnotesize {\mbox{\boldmath $q$}}}\cdot
{\footnotesize {\mbox{\boldmath $r$}}}} {\mathrm d}{\mbox{\boldmath $r$}}
{\mathrm d}{\mbox{\boldmath $r$}}_1 ,
\label{eq:dwia}
\end{equation}
where spin and isospin indices have been omitted for simplicity.

In Eq.~(\ref{eq:dwia})
\begin{equation}
\chi^{(-)}_{E\alpha}({\mbox{\boldmath $r$}}_1) =
\langle\Psi^{\mathrm B}_\alpha(E)|a({\mbox{\boldmath $r$}}_1)|
\Psi_{\mathrm F}\rangle,
\label{eq:dw}
\end{equation}
where $a({\mbox{\boldmath $r$}}_1)$ is an annihilation operator for a nucleon
with coordinate ${\mbox{\boldmath $r$}}_1$, is the s.p. distorted
wave function of the ejectile and the overlap function

\begin{equation}
\left[S_\alpha(E)\right]^{1/2}\phi_{E\alpha}({\mbox{\boldmath $r$}}_1) =
\langle\Psi^{\mathrm B}_\alpha(E)|a({\mbox{\boldmath $r$}}_1)
|\Psi_{\mathrm I}\rangle
\label{eq:ovf}
\end{equation}
describes the residual nucleus as a hole state in the target. The
spectroscopic strength
$S_{\alpha}(E)=\langle \phi_{E\alpha}|\phi_{E\alpha}\rangle $ is the norm
of the overlap integral and gives the probability of removing from the target a
nucleon at ${\mbox{\boldmath $r$}}_1$ leaving the residual nucleus in the state
$|\Psi^{\mathrm B}_{\alpha}(E)\rangle$.

The scattering state in Eq.~(\ref{eq:dw}) and the normalized bound state
$\phi_{E\alpha}({\mbox{\boldmath $r$}}_1)$ in
Eq.~(\ref{eq:ovf}) are consistently derived from an energy-dependent optical
model Feshbach Hamiltonian and, as such, they are not orthogonal. The use of
the effective one-body operator $\hat J^\mu_{\mathrm{eff}}$ in
Eq.~(\ref{eq:dwia}) removes the orthogonality defect of the model wave functions
and takes into account space truncation effects~\cite{BCCGP}. The
orthogonality defect is however negligible in the usual kinematics for the
($e,e'p$) reaction, and in actual DWIA calculations $\hat J^\mu_{\mathrm{eff}}$
is generally replaced by the bare nuclear electromagnetic current operator.
Two-body currents were also included in different theoretical approaches,
but they do not give a relevant contribution to the calculated cross
sections~\cite{BR,VdS,ALC}.

In standard DWIA calculations the nucleon scattering state is eigenfunction of
a phenomenological local spin--dependent optical potential, determined through a
fit to elastic nucleon-nucleus scattering data including cross sections and
polarizations. The nonlocality of the original Feshbach Hamiltonian is taken
into account in the distorted wave function by means of the Perey
factor~\cite{Perey}.

In previous calculations phenomenological bound--state wave functions were usually
adopted for the overlap functions. In the analysis of data these functions
were calculated in a Woods-Saxon well, where the radius and the spectroscopic
factor were considered as free parameters and were determined to reproduce the
experimental momentum distributions, and the well depth was adjusted to
reproduce the experimentally observed separation energy of the bound final
state.

In this paper we have used overlap functions deduced from various correlation
methods which do not contain any free parameters. A short explanation
about how these overlap functions have been calculated on the basis of the OBDM
and their main features are given in sec.~2.2.

The Coulomb distortion of the electron wave functions has been treated with a
high-energy expansion in inverse powers of the electron energy~\cite{DWEEPY}.
For a light nucleus as $^{16}$O, however, an accurate description of Coulomb
distortion is already given by the simple effective momentum approximation,
where the momenta of incoming and outgoing electrons are changed into
effective momenta~\cite{DWEEPY}.

Theoretical results and experimental data are usually presented, for specific
values of the missing energy
$E_{\mathrm m} = \omega-T'-T_{\mathrm B}= E_{\mathrm s} + E_{\mathrm x}$, where
$T'$ and $T_{\mathrm B}$ are the kinetic energies of the outgoing nucleon and
of the residual nucleus, respectively, $E_{\mathrm s}$ is the nucleon
separation energy at threshold and $E_{\mathrm x}$ is the excitation energy of
the residual nucleus, in terms of reduced cross sections, defined by

\begin{equation}
\rho(p_{m})= \int_{\Delta  E_{\mathrm m}}
\frac{{\mathrm d}^{3}\sigma}{{\mathrm d}E'_{0}{\mathrm d}\Omega'_{0}
{\mathrm d}\Omega'} \, \frac{1}{K \sigma_{\mathrm{eN}}} \,
{\mathrm d} E_{\mathrm m},
\label{eq:red}
\end{equation}
as a function of the missing momentum $p_{m}=|{\mbox{\boldmath $p$}}_{\mathrm m}|$,
which is the magnitude of the recoil momentum of the residual nucleus
${\mbox{\boldmath $p$}}_{\mathrm B}$. In Eq.~(\ref{eq:red}),
$K = \Omega_{\mathrm f} f_{\mathrm{rec}}$ and $\sigma_{\mathrm{eN}}$ is the
elementary off-shell electron-nucleon scattering cross section. The CC1
prescription of ref.~\cite{deF} is usually taken for $\sigma_{\mathrm{eN}}$.

Thus, the information contained in the differential cross section is reduced to
a twofold function of $E_{\mathrm m}$ and $p_{\mathrm m}$.
The integral in Eq.~(\ref{eq:red}) is taken over the energy
interval $\Delta E_{\mathrm m}$ that contains the peak of the transition under
study. Calculations are usually performed for a single kinematics,
corresponding to a central value of the phase-space volume in the region of the
peak. Then, the reduced cross section in Eq.~(\ref{eq:red}) can be written
as~\cite{Oxford}
\begin{equation}
\rho^{\mathrm{th}}(p_{\mathrm m})= S_\alpha
n_\alpha(p_{\mathrm m}),
\label{eq:redth}
\end{equation}
where it is assumed that $n_\alpha(p_{\mathrm m})$ is
independent of the energy in the interval $\Delta E_{\mathrm m}$ and
\begin{equation}
S_\alpha = \int_{\Delta E_{\mathrm m}} S_\alpha (E_{\mathrm m})
{\mathrm d} E_{\mathrm m}
\end{equation}
is the spectroscopic factor of the involved hole.

In the plane wave impulse approximation (PWIA), where FSI are neglected,
${\mbox{\boldmath $p$}}_{\mathrm m}$ is equal to the opposite of the initial
momentum of the emitted nucleon in the nucleus, the cross section is factorized as

\begin{equation}
\frac{{\mathrm d}^{3}\sigma}{{\mathrm d}E'_{0}{\mathrm d}\Omega'_{0}
{\mathrm d}\Omega'}= K \sigma_{\mathrm{eN}} \, S_\alpha
\sum_{\alpha}
|\phi_{\alpha}({\mbox{\boldmath $p$}}_{\mathrm m})|^2
\label{eq:pwia}
\end{equation}

and the reduced cross section is the squared Fourier transform of the overlap
function of Eq.~(\ref{eq:ovf}).

In DWIA, the distortion of the electron and outgoing proton waves destroys the
factorization, but it is still useful to define the reduced cross section of
Eq.~(\ref{eq:red}), that can be regarded as the nucleon momentum distribution
modified by distortion and kinematics. This is the quantity that is presented
is Section~3.1.


\subsection{The overlap functions and their relationship with the one-body
density matrix}

The quantities related to the $(A-1)$-particle system, such as the overlap
functions, the separation energies and the spectroscopic factors for its bound
states can be fully determined in principle by the one-body density matrix for the
ground state of the $A$-particle system~\cite{Vn93}. This unique relationship holds
generally for quantum many-body systems with sufficiently short-range forces
between the particles. Below we will give briefly how the overlap function can
be obtained by using the asymptotic behaviour of the OBDM.

The overlap function is related to the hole spectral density function which
takes part in the expression for the cross section of the exclusive
($e,e'N$)
reaction. When the final nuclear state is undetermined, the integral of the
spectral density function over the energy spectrum gives the one-body density
matrix $\rho ({\mbox{\boldmath $r$}},{\mbox{\boldmath $r$}}^{\prime})$.
The latter can be expressed in terms of the overlap functions (\ref{eq:ovf})
in the form:
\begin{equation}
\rho ({\mbox{\boldmath $r$}},{\mbox{\boldmath $r$}}^{\prime})=
\sum_{\alpha }S_{\alpha}\phi_{\alpha}^{*}
({\mbox{\boldmath $r$}})\phi_{\alpha}({\mbox{\boldmath $r$}}^{\prime}).
\label{eq:4}
\end{equation}

In the case of a target nucleus with $J^{\pi }=0^{+}$, each
eigenstate of the residual nucleus is characterized by the
quantum numbers $lj$, i.e., $\alpha \equiv nlj$,
with $n$ being the number of the state with given $l$ and $j$. It is
known~\cite{Mah91} that the overlap functions associated with the bound states
of the $(A-1)$- and $(A+1)$-nucleon systems are eigenstates of a s.p.
Schr\"{o}dinger equation in which the mass operator plays the role of a
potential. Due to its finite range, the asymptotic behaviour of the radial part
of the neutron overlap functions for the bound states of the $(A-1)$--system
is given by~\cite{Vn93,Ber65,Bang85}:
\begin{equation}
\phi_{nlj}(r)\rightarrow C_{nlj}\exp(-k_{nlj}r)/r,
\label{17}
\end{equation}
where

\begin{equation}
k_{nlj}=\frac{1}{\hbar} [2m_{\mathrm n}(E_{nlj}^{A-1}-E_{0}^{A})]^{1/2}.
\label{18}
\end{equation}
In Eq.(\ref{18}) $m_{\mathrm n}$ is the neutron mass, $E_{0}^{A}$ is the
ground state energy of the target $A$-nucleus and $E_{nlj}^{A-1}$ is the
energy of the $nlj$-state of the $(A-1)$--nucleus. For protons some
mathematical complications arise due to an additional long-range
part originating from the Coulomb interaction, but the general conclusions of
the consideration remain the same. The asymptotic behaviour of the radial part
of the corresponding proton overlap functions reads
\begin{equation}
\phi_{nlj}(r)\rightarrow C_{nlj}\exp[-k_{nlj}
r-\eta \ln (2k_{nlj}r)]/r,
\label{11}
\end{equation}
where $\eta $ is the Coulomb (or Sommerfeld) parameter and $k_{nlj}$ (\ref{18})
contains the mass of the proton.

The lowest $(n_{0}lj)$ bound state overlap function (for neutrons) is
determined by the asymptotic behaviour of the corresponding partial radial contribution
of the one-body density matrix  $\rho_{lj}(r,r^{\prime})$ ($r^{\prime}=a\rightarrow \infty
$):
\begin{equation}
\phi _{n_{0}lj}(r)={\frac{{\rho _{lj}(r,a)}}{{C_{n_{0}lj}~\exp
(-k_{n_{0}lj}\,a})/a}}~,
\label{eq:5}
\end{equation}
where the constants ${C_{n_{0}lj}}$ and ${k_{n_{0}lj}}$ are completely
determined by $\rho_{lj}(a,a)$. In this way the separation energy

\begin{equation}
\epsilon _{n_{0}lj}\equiv E_{n_{0}lj}^{A-1}~-~E_{0}^{A}=\frac{\hbar
^{2}~k_{n_{0}lj}^{2}}{2m_{\mathrm n}}~
\label{eq:6}
\end{equation}
and the spectroscopic factor $S_{n_{0}lj}=\langle \phi _{n_{0}lj}\mid \phi
_{n_{0}lj}\rangle $ can be determined as well. As shown in ref.~\cite{Vn93},
the procedure also yields in principle all bound state overlap functions with the same
multipolarity, if they exist. For instance, the overlap function for the next
bound state is:
\begin{equation}
\phi _{n_{1}lj}(r)=\frac{{\rho _{lj}(r,a)}-\phi _{n_{0}lj}(r)
\phi _{n_{0}lj}(a)}{C_{n_{1}lj}~\exp(-k_{n_{1}lj}\,a)/a}.
\label{23}
\end{equation}

The applicability of this theoretical scheme has been demonstrated
in refs.~\cite{Sto96,Vn97,Vn96}. A simple but effective approach accounting for
the SRC within the Jastrow correlation method in its low--order approximation has
been used to obtain the OBDM~\cite{Sto93} and to calculate the s.p. overlap functions
\cite{Sto96}. Another type of overlap functions has been obtained~\cite{Vn97} within the
framework of the CBF theory using the cluster expansion and correlation
functions from Variational Monte Carlo calculations with the Argonne potential.
In refs.~\cite{Gai98,Gai99a} calculations of overlap functions corresponding to
the OBDM~\cite{Sa96} obtained within a model treating correlations independent
on the isospin up to the first order in the cluster expansion and using the
Fermi Hypernetted Chain Technique have been performed. Finally, overlap
functions have been obtained~\cite{Gai98,Gai99a} on the basis of the realistic
OBDM constructed within the GFM~\cite{Po96}. Using the systematic analysis of s.p.
overlap functions obtained from realistic density matrices just mentioned one can
distinguish between the effects of different types of correlations on
quantities such as overlap functions and spectroscopic factors of quasihole
states. In the present paper all these overlap functions are explored for the
analysis of the quasifree nucleon knockout from the $^{16}$O nucleus.

It has been shown in refs.~\cite{Gai98,Gai99a} that due to the inclusion of
short-range as well as tensor correlations the overlap functions are peaked at
smaller distance
in the interior region of the nucleus in comparison with the Hartree-Fock wave
functions. In the momentum space this leads to a slight redistribution of the
strength from the low- to the high-momentum region. Considering the role of
both central and tensor correlations it is found that the correlation effects
on the spectroscopic factors of the hole states are dominated by the tensor
channel of the interaction.

The s.p. overlap functions have been used also as form factors for the
description of $^{16}$O($p,d$)~\cite{Di97,Gai98,Gai99a} and
$^{40}$Ca($p,d$) \cite{Di97} pickup reactions. It was shown, as an important
result from the analysis, that it is not necessary
the theoretically obtained angular distributions to be normalized by means of
spectroscopic factors because the latter are already included in the overlap
functions. Thus having the procedure for calculating such important quantities
as the overlap functions and the spectroscopic factors it is desirable
to apply them also for a consistent study of one-nucleon knockout reactions
induced by electrons and photons, which is the aim of the present paper.


\subsection{The theoretical framework for the ($\gamma,p$) reaction}

The differential cross section for the reaction induced by a photon, with
energy $E_\gamma$, where a nucleon, with momentum ${\mbox{\boldmath $p$}}'$, is
ejected from a nucleus, can be written as~\cite{Oxford}
\begin{equation}
\frac{{\mathrm d}\sigma}{{\mathrm d}\Omega'}= \frac{\pi e^2}{2 E_\gamma}
\,\Omega_{\mathrm f} f_{\mathrm{rec}} W_{\mathrm T},
\label{eq:gp}
\end{equation}
where $\Omega_{\mathrm f}$, $f_{\mathrm{rec}}$ and $W_{\mathrm T}$ have the same
expressions as in Eq.~(\ref{eq:cs}). In contrast to the case of the
electron-induced reaction, where both longitudinal and transverse components of
the nuclear response contribute, here only the pure transverse response
function $W_{\mathrm T}$ occurs.

The DWIA treatment presented in Section~2.1 can be applied also to ($\gamma,N$)
reactions.  Photon-induced nucleon emission appears of particular interest for
our purposes, since the cross sections are expected to be very sensitive to
details of the overlap functions and to effects of SRC. In fact, in this case,
$\omega=|{\mbox{\boldmath $q$}}|=E_\gamma$, and
the mismatch between the momentum transfer and the momentum of the outgoing
nucleon is quite large. Thus, if the reaction proceeds through a DKO mechanism,
only the high-momentum components of the nuclear wave function are probed. On
the other hand, the validity of the DKO mechanism, which is clearly stated for
($e,e'p$), is much more questionable for ($\gamma,N$) reactions, where
two-nucleon processes, such as those involving meson-exchange currents
are expected to be important or even dominant. They are certainly
dominant in the ($\gamma,n$) reaction, where the DKO mechanism, even in its
most sophisticated version~\cite{BCGP}, gives but a small fraction of the
experimental cross sections, while for the ($\gamma,p$) reaction the
contribution of DKO is much more relevant.
Nonrelativistic DWIA calculations, based on the same approach presented in
Section~2.1 for the ($e,e'N$) reaction~\cite{BGP}, and more recent
relativistic calculations, also based on the DKO mechanism~\cite{relativ}, are
able to give a fair description of data. The results, however, are very
sensitive to the theoretical ingredients adopted for bound and scattering
states. On the other hand, various calculations in different theoretical
approaches indicate that MEC play a prominent role also in the ($\gamma,p$)
reaction \cite{Benenti,MEC}.

Therefore in this paper the cross sections of the exclusive ($\gamma,p$)
reaction have been  calculated in the theoretical framework of
ref.~\cite{Benenti}, where photoabsorption occurs, through one-body and
two-body currents, on a pair of correlated nucleons: only one of them is then
emitted, while the other one is reabsorbed in the residual nucleus. In this
model MEC are included in the framework of the DKO model with FSI and thus the
size and the relative weight of DKO and MEC can be evaluated consistently.

In the calculations of ref.~\cite{Benenti}  the correlated wave function of
the pair was given by the product of shell-model s.p. bound
state wave functions and of a central Jastrow type correlation function,
which takes into account SRC. Here we have adopted one-nucleon overlap
functions obtained, as it is explained in Section~2.2, from different realistic
one-body density matrices and that already include SRC. The calculated
cross section is thus the sum of two terms: the direct contribution of the
one-body current, which corresponds to the quasifree DKO considered in
Section~2.1, and the exchange contribution of the two-body current. In this
model MEC are explicitly taken into account in a microscopic and unfactorized
calculation. Moreover, a consistent analysis, with consistent ingredients, i.e.
overlap functions, spectroscopic factors and optical model parameters, can be
performed of ($e,e'p$) and ($\gamma,p$) reaction cross sections.

In order to reduce the complexity of the calculation, we have adopted the same
approximations as in ref.~\cite{Benenti}. Only the contribution of the
two-body current due to the seagull diagrams has been included. Currents due to
the pion-in-flight diagrams give a pure contribution much smaller than that
due to the seagull diagrams and currents corresponding to diagrams with
intermediate $\Delta$ isobar configurations become important only for photon
energies above the pion-production threshold. Thus, the seagull current, here
considered, should give the main contribution of the two-body current in the
photon-energy range above the giant resonance and below the pion production
threshold.

The spin-orbit part of the optical potential has been neglected in the
calculations of the ($\gamma,p$) cross section, as well as various effects
considered in ref.~\cite{BCGP} in the framework of the DKO model:
charge-exchange FSI, orthogonality between initial and final states,
antisymmetrization of the outgoing particle and recoil terms. These effects
cannot be simply applied to the present approach, where also two-body currents
are included. Their evaluation would require a specific and consistent
treatment. In ref.~\cite{BCGP} the sum of these effects turns out to be very
important for ({$\gamma,n$), while for ($\gamma,p$) it  gives a much smaller
contribution, which should be further reduced when also the two-body current is
added in the calculated cross sections.

Thus, although with some approximations, the present treatment should include
all the most important and essential ingredients contributing to the cross
section of the ($\gamma,p$) reaction in the photon-energy range between 50 and
100 MeV.


\section{Results and discussion}
\subsection{The $^{16}$O($e,e'n$) and $^{16}$O($e,e'p$) reactions}

In this section we discuss results for the reduced cross sections of the
$^{16}$O($e,e'n$) and $^{16}$O($e,e'p$) reactions. In PWIA, where the effects
of the FSI between the outgoing nucleon and the residual $(A-1)$-nuclear system
are neglected, the reduced cross section $\rho(p_{\mathrm m})$, defined in
Eq.~(\ref{eq:redth}), is the squared Fourier transform of the overlap funtion.
In DWIA it can be regarded as the nucleon momentum distribution modified by
distortion and kinematics.

In standard DWIA calculations the overlap function is generally replaced by
a phenomenological s.p. bound--state wave function which is eigenfunction of a
mean--field Woods-Saxon potential. In the analysis of the ($e,e'p$) data the
well depth is determined to reproduce the separation energy values and the
radius is adjusted to fit the shape of the momentum distribution. In the
present work we have used overlap functions derived from different calculations
of the OBDM. We would like to emphasize that this is the correct theoretical
procedure which in principle has to be used for an accurate description of
($e,e'N$) knockout reactions. Our calculations are performed by using the s.p.
overlap functions obtained from Eqs.~(\ref{17}) and (\ref{11}) for the neutron
and proton bound states, respectively. The results presented for both
($e,e'n$) and ($e,e'p$) reactions are obtained using the neutron overlap
functions. The application of the latter to the ($e,e^{\prime}p$) reaction can be
justified by the fact that the proton and neutron overlap functions are very similar
and the use of the correct proton one leads to almost identical results in this case.

Our analysis is made for the transitions to the $1/2^{-}$ ground state and to
the first $3/2^{-}$ excited state of the residual nucleus (at excitation energy
$E_{\mathrm x}=6.18$ MeV for $^{15}$O in the case of the ($e,e'n$) reaction
and at $E_{\mathrm x}=6.3$ MeV for $^{15}$N in the case of the ($e,e'p$)
reaction), representing a knockout from the valence $1p$ shell of $^{16}$O.
Unfortunately, experiments on the ($e,e'n$) reactions have not been
carried out so far, due to the difficulties in performing neutron detection
in a coincidence reaction and, therefore, our results are theoretical predictions.
In contrast, a large number of experiments have been performed over the past years on
the ($e,e'p$) reactions~\cite{FM84,Oxford,Lap93,Leuschner,Kra90,Steen88,Her88}. We
compare here our theoretical results for the $^{16}$O($e,e'p$) reaction with the data
taken at NIKHEF~\cite{Leuschner} in the so-called parallel kinematics. In this
kinematics the momentum of the outgoing nucleon is fixed and is taken
parallel or antiparallel to the momentum transfer. Different values of the
missing momentum are obtained by varying the electron scattering angle
and therefore the magnitude of the momentum transfer. This kinematics, that
has been considered in most of the ($e,e'p$) experiments, represents a
suitable example for the present investigation of the effects of different
overlap functions in quasifree nucleon knockout reactions from $^{16}$O.

The reduced cross sections for the $^{16}$O($e,e'n$) reaction as a function of
the missing momentum and for the transitions to the $1/2^{-}$ ground state and
the $3/2^{-}$ excited state of $^{15}$O are displayed in Fig.~1. The results
obtained with different overlap functions are compared with those given by the
Hartree-Fock (HF) wave function, which is calculated in a self-consistent way
using the Skyrme-III interaction. Besides the HF wave function, whose norm is
equal to one, all the overlap functions contain a spectroscopic factor. These
factors are listed in Table I (column I) and were discussed in details
in refs.~\cite{Gai98,Gai99a}. They account for the contribution of correlations
included in the OBDM which cause a depletion of the quasihole states. Only
short--range central correlations are included in the OBDM of
refs.~\cite{Sto96,Sa96}, whereas also tensor correlations are taken into
account in refs.~\cite{Vn97,Po96}. It was found that correlation effects on the
spectroscopic factor of the hole states are dominated by the tensor channel of
the interaction~\cite{Gai98,Gai99a}. Indeed the spectroscopic factors in Table
I are lower for the overlap functions including also tensor correlations.
These overlap functions, however, do not include LRC, which
should produce further depletion of the quasihole states~\cite{Amir,Geurts}.

The reduced cross sections in Fig.~1 are sensitive to the shape of the
various overlap functions used. The differences  are considerable at large
values of $p_{\mathrm m}$, where the cross section is several orders of
magnitude lower than in the maximum region. Similar results for the different
overlap functions as a function of $p_{\mathrm m}$ have been obtained also for
the ($e,e'p$) reaction and in different kinematics. The deviations of the
various results at large values of $p_{\mathrm m}$ are related to different
accounting for the short-range NN correlations within the correlation methods
used. SRC are particularly important in one-nucleon emission at large missing
momenta and energy~\cite{PR,MD}. At high missing energies, however, other
competing processes are also present and a clear identification of SRC can
better be made by means of two-nucleon knockout reactions~\cite{tnko}. At low
missing-energy values measurements over an extended range of missing momenta,
in particular at large values, where the SRC effects seem to be more sizable,
can test the various s.p. overlap functions and NN correlations.

The reduced cross sections for the $^{16}$O($e,e'p$) reaction are shown in
Fig.~2 in comparison with data. In order to reproduce the size of the
experimental cross section a reduction factor has been applied to the
theoretical results. These factors, which have been obtained by a fit of the
calculated reduced cross sections to the data over the whole missing-momentum
range considered in the experiment, are also listed in Table I (column II). In
general, a good agreement with the shape of the experimental distribution is
achieved. The results, however, are also sensitive to details of the various
overlap functions. The best agreement with the data, for both transitions, is
obtained with the overlap functions~\cite{Gai98,Gai99a} emerging from the OBDM
calculated within the Green function method~\cite{Po96}.
This is due to the substantial realistic inclusion of short-range as well as
tensor correlations in the OBDM. The calculations based on the Green function
theory~\cite{Gai98,Gai99a} have shown that about 10 $\%$ of the $1p$ strength
is removed by these correlations. The reduced cross section obtained by using
the overlap function from ref.~\cite{Vn97} gives also a good agreement for the
$1/2^{-}$  state, but a less satisfactory description of data for $3/2^{-}$. In
contrast, the overlap function from ref.~\cite{Sto96} gives a better
description of the experimental distribution for the $3/2^{-}$ than for the
ground state. The shape of the experimental reduced cross sections can
adequately be described also by the HF wave functions, in particular for
$p_{\mathrm m}\leq 150$ MeV/$c$. Only the overlap function extracted from the
OBDM of ref.~\cite{Sa96} is unable to give an adequate description of the
experimental momentum distributions of the $1/2^{-}$ state, while it gives a
better agreement for the $3/2^{-}$ state. In general, the agreement of the
calculated reduced cross sections with data is somewhat better for the
$3/2^{-}$ than for the $1/2^{-}$ state.

We note that even though a fair agreement with the shape of the experimental
distributions is generally obtained in the present calculations, this agreement
is not as good as in the analysis of ref.~\cite{Leuschner}. The calculations
have been performed in both cases with the same DWIA treatment and with the
same optical potential \cite{Schwandt}, but in ref.~\cite{Leuschner} a s.p. phenomenological
wave function was adopted, with some parameters adjusted to the data. In the present
work overlap functions obtained within different correlation methods, which contain
approximations but no free parameters have been used.

Only a reduction factor has been applied to the calculated cross sections to
reproduce the data. The fact that our results overestimate the data may be
explained on a theoretical basis by the observation that our overlap functions
are deduced from calculations including only SRC but not LRC. The reduction
factor can thus be considered as a further spectroscopic factor reflecting the
depletion of the quasihole state produced by LRC. Of course, the discrepancy
with the data can be due also to other effects not included or not adequately
described by the theoretical treatment. For instance, a relativistic optical potential
increases by about 15\% the absorption due to FSI and thus gives a reduction of the
calculated cross sections~\cite{relat,darwin}. On the other hand, a proper treatment of the
center-of-mass motion leads to an enhancement of the spectroscopic factor by
about 7\%~\cite{CM}. Also two-body currents may lead to small variations of the
size of the calculated cross sections~\cite{VdS}. We note, however, that the
reduction factors applied here to the calculated cross sections are not the
result of a precise theoretical calculation. They have been obtained by a fit
to the data and have only an indicative meaning. Small variations within
10-15\% around their values would not significantly change the comparison with
data. In any case the reduction factors should mostly be ascribed to LRC, but
for the HF wave function, which does not contain any kind of correlations. For
this wave function the reduction factor accounts for both LRC and SRC. It is
interesting to note that in the calculations with all the correlated overlap
functions the reduction factors for the $1p_{1/2}$ state turn out to be close
to the spectroscopic factor (0.83) obtained in the theoretical approach of
ref.~\cite{Amir} where only LRC are included. For $1p_{3/2}$, however, the
reduction factors are lower than the spectroscopic factor (0.85) calculated in
the same approach.

In Table I we give, in addition, in column III the factor obtained by the
product of the two factors in columns I and II. This factor can be considered
as a total spectroscopic factor and can be attributed to the combined effect
of SRC and LRC. Indeed for $1p_{1/2}$ these factors are in reasonable agreement
with the spectroscopic factor (0.76) calculated in ref.~\cite{Geurts}, where
both SRC and LRC are consistently included. Also the HF wave function gives a
total spectroscopic factor in agreement with the result of ref.~\cite{Geurts}
and a reasonable description of the shape of the experimental distribution in
Fig.~2. This means that in the missing--momentum range considered by the
experiment, the correlation effects  are overwhelmed by the dominant quasihole
component already present in the HF approximation. For the
transition to the $1/2^{-}$ state a quite large value of the total
spectroscopic factor is obtained with the overlap function extracted from the OBDM by
adopting the Average Correlation Approximation~\cite{Sa96}. This is due to the fact
that in this approach the correlations are mainly produced by the central short--range
components of the NN interaction. Moreover, this function is unable to reproduce correctly
the shape of the experimental distribution.

The total spectroscopic factors obtained for the $3/2^{-}$ state are lower than
those calculated in ref.~\cite {Geurts}, but for the overlap function from
ref.~\cite{Vn97}, which, on the other hand, gives for this state a worse
description of the data. We note that also other analyses of the same
data~\cite{Leuschner,PR,Amir} gave for $3/2^{-}$ a spectroscopic factor lower
than for the ground state. It was noticed in ref.~\cite{PR} that three
$3/2^{-}$ states are observed in $^{15}$N at low excitation energies and that
LRC yield a splitting such that 86\% of the total strength going to these
states is contained in the data. This splitting is not observed in the
calculations. If the total spectroscopic factors in Table I are divided by
0.86 to account for the splitting of the experimental strength, we obtain a
value closer to that obtained for the $1/2^{-}$ state and in the calculation of
ref~\cite{Geurts}.

In the DWIA analysis of the data in ref.~\cite{Leuschner} phenomenological
Woods-Saxon bound-state wave functions gave the "experimental" spectroscopic
factors of 0.61 for $1/2^{-}$ and 0.53 for $3/2^{-}$, somewhat lower than in
the present work with theoretically calculated overlap functions.

The overlap functions here considered only include SRC and some of them also
tensor correlations. It would be of great interest to avail theoretically
calculated overlap functions where both SRC and LRC are consistently included
and to investigate their effects on the size and the shape of the calculated
cross sections and in comparison with data.


\subsection{The $^{16}$O($\gamma,p$) reaction}

In this section we present results of calculations for the
$^{16}$O($\gamma,p$) reaction and discuss them in comparison with data.
Calculations have been performed within the theoretical framework of
ref.~\cite{Benenti}, where one-body and two-body currents are included and both
contributions of DKO and MEC can be evaluated consistently. The same
theoretical ingredients, i.e. s.p. overlap functions, spectroscopic factors and
consistent optical potentials, have been adopted as in the calculations of the
($e,e'p$) cross section. Moreover, the reduction factor determined in
comparison with the ($e,e'p$) data has been applied also in the comparison of
the calculated ($\gamma,p$) cross section with data. Since ($\gamma,p$)
calculations are extremely sensitive to the theoretical ingredients adopted for
bound and scattering states, the use of constrained parameters should allow us
to reduce ambiguities in the interpretation of the results and to perform a
consistent study of the $^{16}$O($e,e'p$) and $^{16}$O($\gamma,p$) reactions.

The aim of our investigation is twofold. On the one hand, we intend to check
the consistency of the theoretical treatment for the two reactions. On the
other hand, we want to investigate the sensitivity of the results to the
various overlap functions and to NN correlations, which are included in the
overlap functions within different theoretical frameworks, at large values of
the missing momentum, where SRC effects are more sizable.

We restrict our analysis to photon-energy values where our theoretical
treatment appears more reliable. Therefore we have performed calculations at
$E_\gamma=60$ and $72$ MeV, where $^{16}$O($\gamma,p$) data are available for
the transition to the $1/2^{-}$ ground state~\cite{FO,Smet,Miller} and to the
$3/2^{-}$ excited state at 6.3 MeV~\cite{Miller}. At these photon-energy values
it is possible to sample in comparison with data $p_{\mathrm m}$ values between
250 and 400 MeV/$c$.

The angular distribution of the $^{16}$O($\gamma,p$)$^{15}$N$_{\mathrm{g.s.}}$
reaction at $E_\gamma = 60$  MeV is displayed in Fig.~3. In the figure the
results given by the sum of the one-body and of the two-body seagull currents
are compared with the contribution given by the one-body current, which roughly
corresponds to the DWIA treatment based on the DKO mechanism.

The DWIA calculations with different overlap functions exhibit considerable
differences, in particular at backward angles, where larger values of
$p_{\mathrm m}$ are probed. These differences are somewhat reduced when the
two-body seagull current is added, but remain anyhow quite large.

The contribution of the one-body current represents a large part of the measured
cross section, but none of the overlap functions used is able to give a proper
description of the data. All the curves in DWIA lie well below the data, but
that obtained with the overlap function from the OBDM of ref.~\cite{Sa96}, which
is anyhow able to reproduce the size of the experimental cross section only at
the lowest angles. A much better agreement with ($\gamma,p$) data is obtained
when MEC are added to the DWIA result. The HF wave function and the overlap function
obtained from the OBDM of ref.~\cite{Sto96} are able to reproduce the size of the
experimental cross section, but only at low values of the outgoing proton
angle. Thus they are unable to reproduce the shape of the distribution. The result with
the overlap function from the OBDM of ref.~\cite{Sa96} largely overshoots the data. Much
better agreement with the shape of the experimental distribution is given by
the other correlated overlap functions. We notice that a better overlap with the
$(\gamma,p)$ data is obtained using the overlap function from the OBDM of ref.~\cite{Po96}
and, to a lesser extent, also by that from ref.~\cite{Vn97}. These overlap
functions, obtained from calculations where short--range as well as tensor
correlations are included, are also able to give the best agreement with
($e,e'p$) data for the $1/2^{-}$ state in Fig.~2. In Fig.~3 calculations with
these wave functions lie a bit below the data in the maximum region, but they
give a fair agreement with the shape of the experimental distribution, in
particular for the overlap function from ref.~\cite{Po96}. The existing
discrepancies are anyhow not large and might be explained within the
approximations of the theoretical model, or also by a bit lower reduction
factor. We already observed in the ($e,e'p$) analysis of Section~3.1 that small
variations around the values listed in Table I would not change significantly
the comparison with ($e,e'p$) data, but would here improve the agreement with
the experimental results for the ($\gamma,p$) reaction, which is much more
sensitive to the various theoretical ingredients.
The fact that the overlap function from the OBDM of ref.~\cite{Po96} is able to
give the best description of both ($e,e'p$) and ($\gamma,p$) data is a strong
indication in favour of a consistent analysis of the two reactions.
Similar results are obtained in Fig.~4, where the angular distribution of the
$^{16}$O($\gamma,p$)$^{15}$N$_{\mathrm{g.s.}}$ reaction is displayed at
$E_\gamma = 72$  MeV. The results confirm the important role played by MEC to
describe the size and the shape of the experimental cross section and the great
sensitivity to the shape of the overlap function and to correlation effects.
The behaviour of the angular distribution in comparison with data is similar to
that of Fig.~3, at $E_\gamma = 60$ MeV. Also at $E_\gamma = 72$ MeV the HF
wave function and the overlap function from the OBDM of ref.~\cite{Sto96} can
reproduce the size of the experimental cross section only at low values of the
scattering angle and are unable to give a proper description of the shape of
the angular distribution. The cross section calculated with the overlap
function from ref.~\cite{Sa96} overshoots the data, while a fair agreement with
data is obtained with the overlap function from ref.~\cite{Po96}. The cross
section calculated with this overlap function as well as that calculated with
the overlap function from ref.~\cite {Vn97} are a bit higher than the data at
low values of the outgoing proton angle, but the discrepancy is about the same
as at $E_\gamma = 60$ MeV and might be explained within the uncertainties of
the theoretical treatment.

An example for the transition to the $3/2^{-}$ state at 6.3 MeV is presented in
Fig.~5, where the angular distribution of the $^{16}$O($\gamma,p$) reaction at
$E_\gamma = 72$ MeV is displayed in comparison with data. Similar results have
been obtained at $E_\gamma = 60$ MeV. The results for this transition confirm
the sensitivity to the overlap function and the important role of MEC in the
cross section of the ($\gamma,p$) reaction. However, the conclusions about
comparison with data are in this case less clear. The size of the experimental
cross section is already described by DWIA calculations, but with the overlap
function derived from the OBDM of ref.~\cite{Sto96}. In contrast, the shape is
much better reproduced by the more complete calculations including also the
seagull current. On the other hand, these results including both one-body and
two-body currents overshoot the data, but with the overlap function from
ref.~\cite{Sto96}. The overlap function from ref.~\cite{Po96}, which is able to
give the best agreement with ($e,e'p$) data also for this transition, is able
to reproduce very well the shape of experimental angular distribution in
Fig.~5, but the calculated cross section overshoots the data by a factor of
about two. Only the overlap function from the OBDM of ref.~\cite{Sto96}
gives a fair agreement with the size and the shape of the ($\gamma,p$) data in
Fig.~5. The small discrepancy might be explained within the approximations of
the model. For instance, an enhancement of the cross section at high values of
the scattering angle should be given by the spin-orbit part of the optical
potential~\cite{BGP}, which has been neglected in the present approach. We want
to remind that the overlap function from ref.~\cite{Sto96} is also able to give
in Fig.~2 a very good description of the ($e,e'p$) experimental reduced cross
section for the transition to the same $3/2^{-}$ state.

The results for the $3/2^{-}$ state, although less clear than for the
ground state of $^{15}$N, can thus be considered as further evidence in favour
of a consistent description, with the same theoretical ingredients, of
($e,e'p$) and ($\gamma,p$) data. However, in order to draw definite
conclusions, a more refined theoretical treatment of the ($\gamma,p$) reaction
is needed, where the approximations of the present approach are improved and a
more careful comparison with data in a wider photon-energy range can be
performed.


\section{Summary and conclusions}

Single--particle overlap functions calculated, for the $^{16}$O nucleus, on the
basis of the OBDM emerging from various correlations methods have been used to
calculate the cross sections of the ($e,e'n$), ($e,e'p$) and ($\gamma,p$)
reactions, for the transitions to the $1/2^{-}$ ground state and the first
$3/2^{-}$ excited state of the residual nucleus. These overlap functions
contain short--range central and tensor correlations and include the
spectroscopic factor. The aim of the present investigation was to clarify the
importance of various types of correlations, which are accounted for to
different extent in the theoretical methods considered, on the reaction cross
sections and in comparison with data.

The reduced cross sections of the $^{16}$O($e,e'n$) and  $^{16}$O($e,e'p$)
knockout reactions have been calculated with the same nonrelativistic DWIA
treatment which was successfully applied previously to the analysis of many ($e,e'p$)
data. In the standard DWIA approach, however, phenomenological s.p. wave functions were
used, with some parameters fitted to the data. In this paper the results have been
obtained with theoretically calculated overlap functions which do not include free
parameters.

The reduced cross sections are sensitive to the shape of the various overlap
functions and exhibit considerable differences at large values of the missing
momentum, where correlation effects are more sizable. The theoretical results
are generally able to reproduce, with a fair agreement, the shape of the
experimental reduced cross sections. The quality of the agreement, however, is
sensitive to details of the different overlap functions.

In order to reproduce the size of the experimental data a reduction
factor must be applied to the calculated reduced cross sections. This factor,
that is extracted from a fit to the data, can be considered as a further
spectroscopic factor to be mostly ascribed to LRC, which also cause a
depletion of the quasihole states and which are not included in the overlap
functions considered here. The spectroscopic factors accounting for SRC and
LRC obtained in the present analysis are in reasonable agreement with those
given by previous theoretical investigations.

Since both SRC and LRC have sizable effects on the spectroscopic factors, on
the shape of the overlap function and, as a consequence, on the cross sections,
a calculation of fully correlated overlap functions consistently including SRC
and LRC, although extremely difficult, would be highly desirable and would
allow a direct and parameter-free comparison with data.

The behaviour of the different overlap functions at high values of momenta and
NN correlation effects can better be investigated in the ($\gamma,p$) reaction.
The cross section of the $^{16}$O($\gamma,p$) reaction has been calculated at
$E_\gamma = 60$ and $72$ MeV. Consistent theoretical ingredients and the same
spectroscopic factors extracted from the analysis of the $^{16}$O($e,e'p$)
reaction have constrained the calculations. The theoretical treatment of the
photon induced reaction includes both contributions of the DKO mechanism and of
the two-body pion seagull current.

The various overlap functions give considerable differences on the size and
shape of the calculated cross sections. The contribution of the DKO mechanism
is unable to describe the experimental data. The numerical results generally
fall short of the data and all the curves are unable to reproduce the shape of
the experimental angular distributions. The contribution of MEC is large. It
significantly affects both the size and shape of the cross section and
generally brings the calculated cross section in much better agreement with the
data.

For the transition to the ground state of $^{15}$N the best and a fair
description of the data is given by the overlap function able to give also the
best description of the ($e,e'p$) data. This is a strong indication in favour
of the consistency in the analysis of the ($e,e'p$) and ($\gamma,p$) reactions.
This result is partly confirmed also for the  transition to the $3/2^{-}$
excited state. In this case, however, the situation is less clear and larger
discrepancies in comparison with data are obtained. These discrepancies are
anyhow not too large and might be explained within the approximations of the
model.

A more refined theoretical treatment, which should consistently include SRC
and LRC, as well as orthogonality, antisymmetry and c.m. effects, together with
a more complete evaluation of two-body currents, would allow a more careful
comparison with data in a wider kinematical range and would be needed to draw
definite conclusions.

\acknowledgments
One of us (C.G.) wishes to thank F.D. Pacati and M. Radici for useful
discussions. The authors thank H. M\"{uther}, G. Co' and D. Van Neck for providing
us the results for the OBDM used. The work is partly supported by
the Bulgarian National Science Foundation under Contract $\Phi$--809.

\newpage


\noindent {\bf Table 1:} Spectroscopic factors for the $^{16}$O($e,e'p$)
knockout reaction leading to the $1/2^{-}$ ground state and to the $3/2^{-}$
excited state of $^{15}$N. Column I gives the spectroscopic factors deduced
from the calculations with different OBDM of $^{16}$O; II gives the additional
reduction factors determined through a comparison between the ($e,e'p$) data of
ref.~\cite{Leuschner} and the reduced cross sections calculated in DWIA with
the different overlap functions; III gives the total spectroscopic factors
obtained from the product of the factors in columns I and II.
\vspace{1cm}

\begin{center}
\begin{tabular}{cccccccc}
\hline\hline
&  \multicolumn{3}{c}{$1p_{1/2}$} & &  \multicolumn{3}{c}{$1p_{3/2}$} \\
\cline{2-4} \cline{6-8}
OBDM & I   & II  & III & & I   & II  & III \\
\hline
HF               & 1.000 & 0.750 & 0.750 & &  1.000 & 0.550 & 0.550\\
JCM \cite{Sto96} & 0.953 & 0.825 & 0.786 & &  0.953 & 0.600 & 0.572\\
CBF \cite{Vn97}  & 0.912 & 0.850 & 0.775 & &  0.909 & 0.780 & 0.709\\
CBF \cite{Sa96}  & 0.981 & 0.900 & 0.883 & &  0.981 & 0.600 & 0.589\\
GFM \cite{Po96}  & 0.905 & 0.800 & 0.724 & &  0.915 & 0.625 & 0.572\\
\hline\hline
\end{tabular}
\end{center}
\newpage


\begin{figure}
\caption[]{Reduced cross section of the $^{16}$O($e,e'n$) reaction as
a function of the missing momentum $p_{\mathrm m}$ for
the transitions to the $1/2^{-}$ ground state and the first $3/2^{-}$ excited state
of $^{15}$O in parallel kinematics, with $E_0=520.6$ MeV and an outgoing neutron
energy of 100 MeV. The optical potential is from ref.~\cite{Schwandt}. Overlap functions
derived from the OBDM of GFM \cite{Po96} (solid lines), CBF \cite{Vn97} (dashed
lines), CBF \cite{Sa96} (dot-dashed lines), JCM \cite{Sto96} (double
dot-dashed lines). The dotted lines are calculated with the HF wave
function. The positive (negative) values of $p_{\mathrm m}$ refer to situations
where $|{\mbox{\boldmath $q$}}|<|{\mbox{\boldmath $p$}}'|$
($|{\mbox{\boldmath $q$}}|>|{\mbox{\boldmath $p$}}'|$).
\label{fig:een}
}
\end{figure}

\begin{figure}
\caption[]{Reduced cross section of the $^{16}$O($e,e'p$) reaction as
a function of the missing momentum $p_{\mathrm m}$ for the transitions to the
$1/2^{-}$ ground state and the first $3/2^{-}$  excited state of $^{15}$N in
parallel kinematics, with $E_0=520.6$ MeV and an outgoing proton energy of 90
MeV. The optical potential is from ref.~\cite{Schwandt} (see Table
III of ref.~\cite{Leuschner}). Line convention is as in Fig.~1. The experimental
data are taken from ref.~\cite{Leuschner}. The theoretical results have been
multiplied by the reduction factor given in column II of Table I.
\label{fig:eep}
}
\end{figure}

\begin{figure}
\caption[]{Angular distribution of the cross section of the
$^{16}$O($\gamma,p$) reaction for the transition to the $1/2^{-}$ ground state
of $^{15}$N  at $E_\gamma = 60$ MeV. The separate contribution given by the
one-body current (DWIA) and the final result given by the sum of the one-body
and the two-body seagull current (DWIA+MEC) are shown. Line convention is as in
Fig.~1. The optical potential is from ref.~\cite{Schwandt}. The experimental data are
taken from ref.~\cite{FO} (black circles), ref.~\cite{Smet} (open circles) and
ref.~\cite{Miller} (triangles). The theoretical results have been multiplied by
the reduction factors listed in column II of Table I, consistently with the
analysis of ($e,e'p$) data.
\label{fig:gp60}
}
\end{figure}

\begin{figure}
\caption[]{Angular distribution of the cross section of the
$^{16}$O($\gamma,p$) reaction for the transition to the $1/2^{-}$ ground state
of $^{15}$N  at $E_\gamma = 72$ MeV. The separate contribution given by the
one-body current (DWIA) and the final result given by the sum of the one-body
and the two-body seagull current (DWIA+MEC) are shown. Line convention is as in
Fig.~3. The optical potential is from ref.~\cite{Schwandt}. The experimental data are
taken from ref.~\cite{Miller}. The theoretical results have been multiplied by
the reduction factors listed in column II of Table I, consistently with the
analysis of ($e,e'p$) data.
\label{fig:gp72}
}
\end{figure}

\begin{figure}
\caption[]{Angular distribution of the cross section of the
$^{16}$O($\gamma,p$) reaction for the transition to the $3/2^{-}$ state of
$^{15}$N at 6.3 MeV and at $E_\gamma = 72$ MeV. The separate contribution
given by the one-body current (DWIA) and the final result given by the sum of
the one-body and the two-body seagull current (DWIA+MEC) are shown. Line
convention and the optical potential are as in Fig.~4. The experimental data are taken
from ref.~\cite{Miller}. The theoretical results have been multiplied by the
reduction factors listed in column II of Table I, consistently with the analysis
of ($e,e'p$) data.
\label{fig:gp72p3}
}
\end{figure}
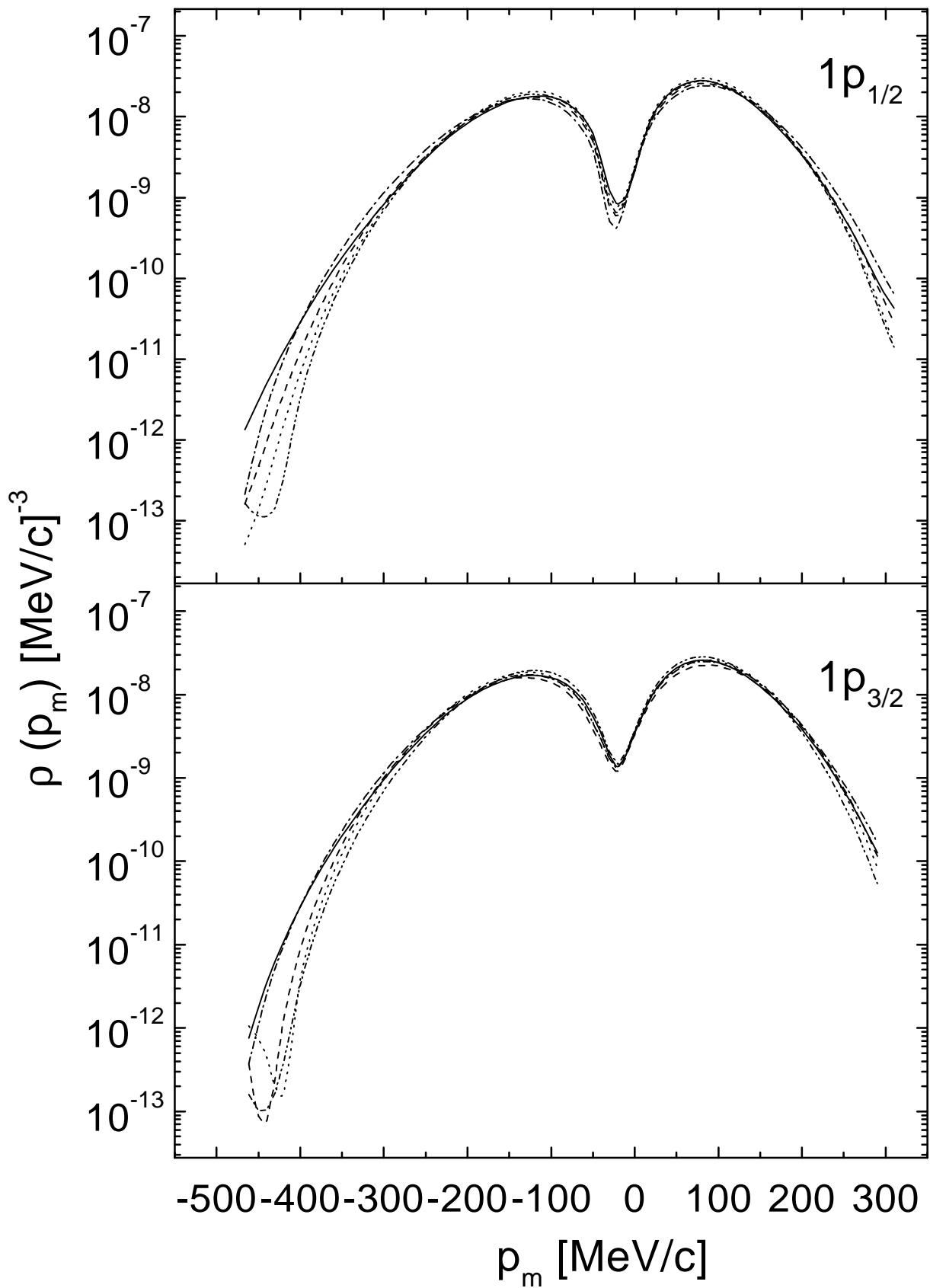
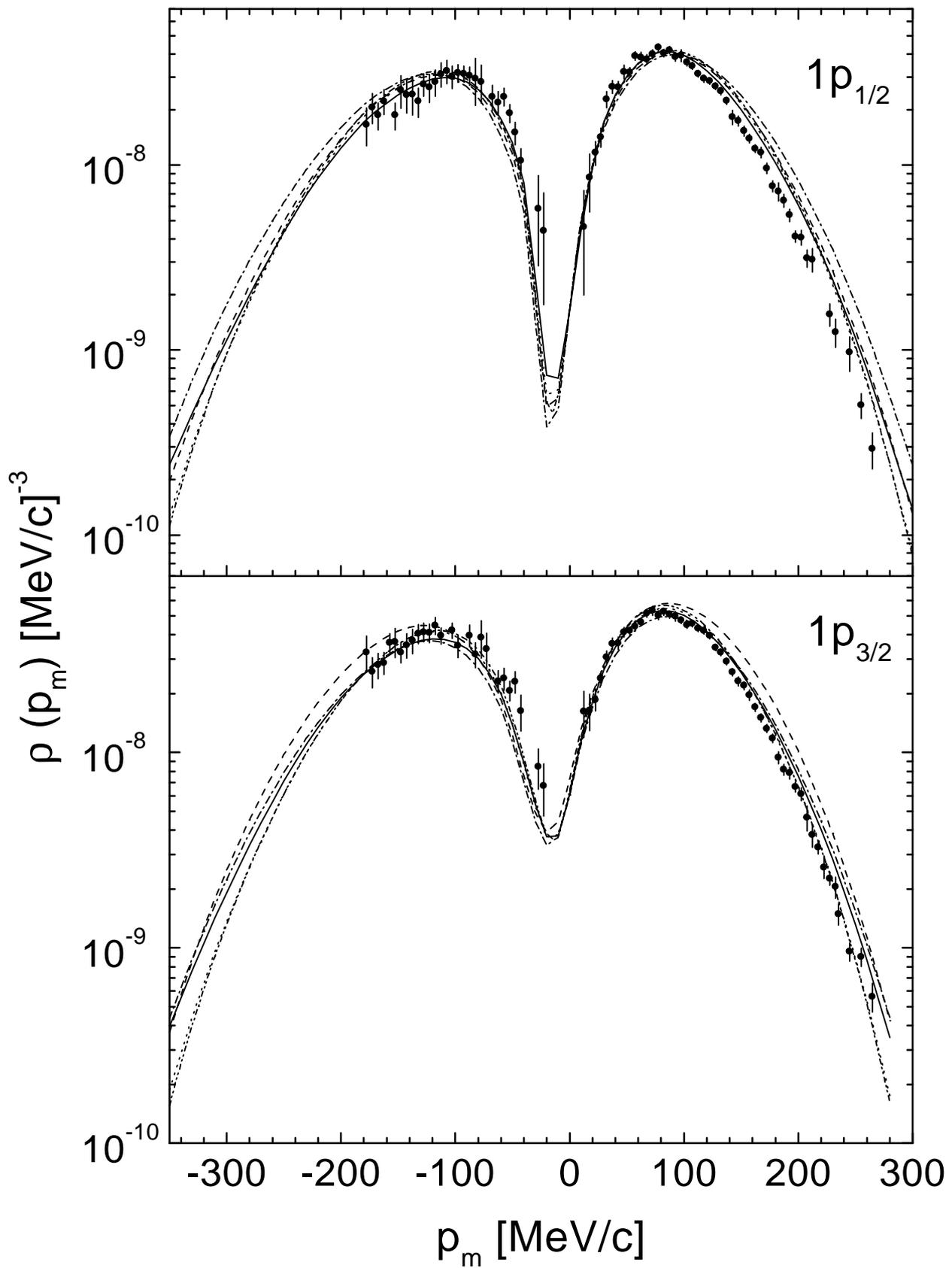
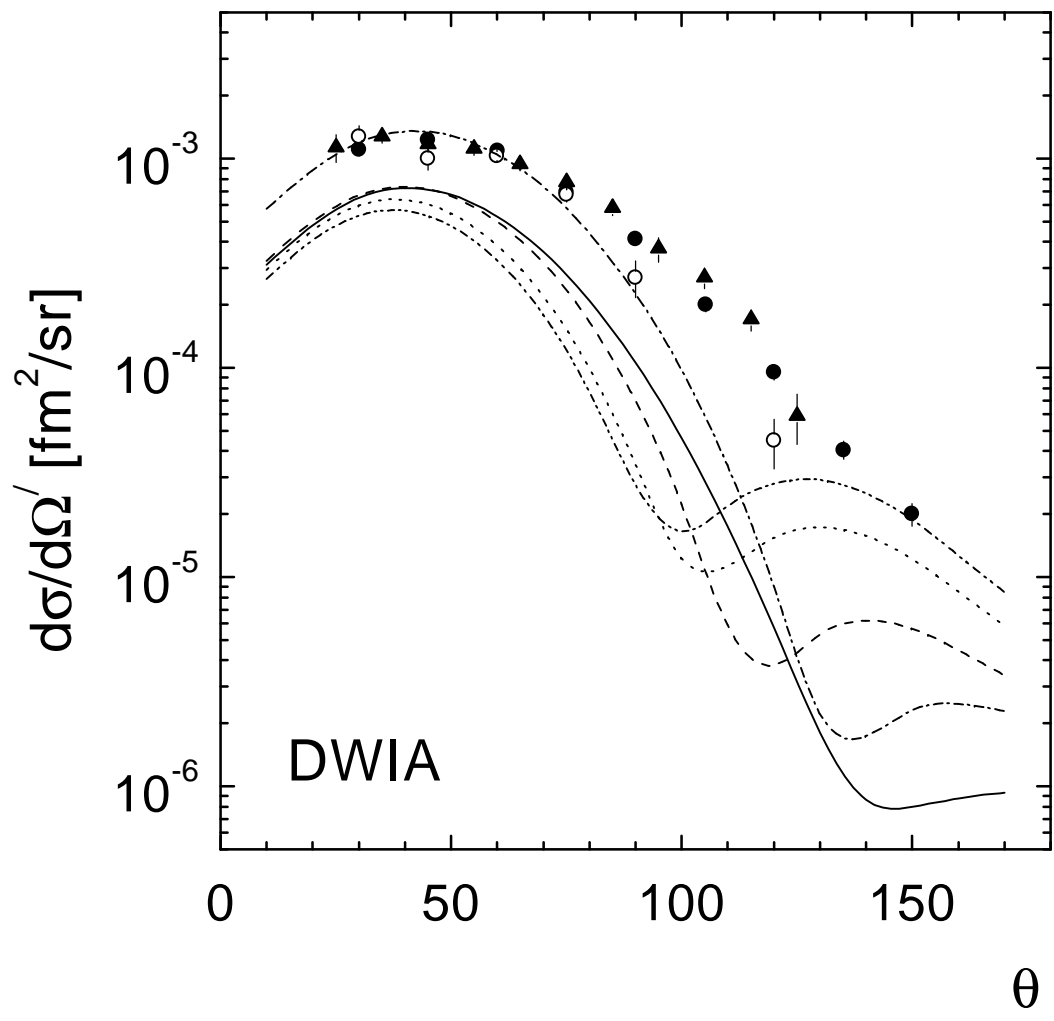
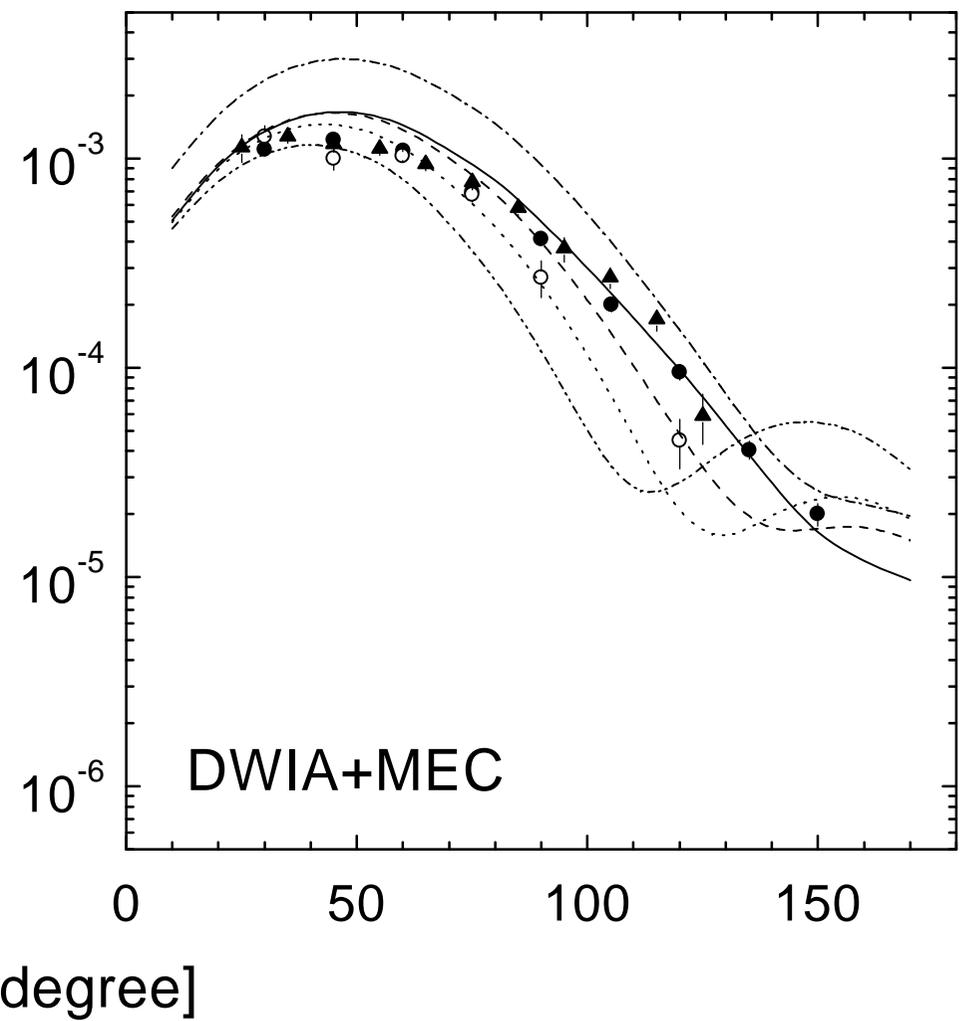
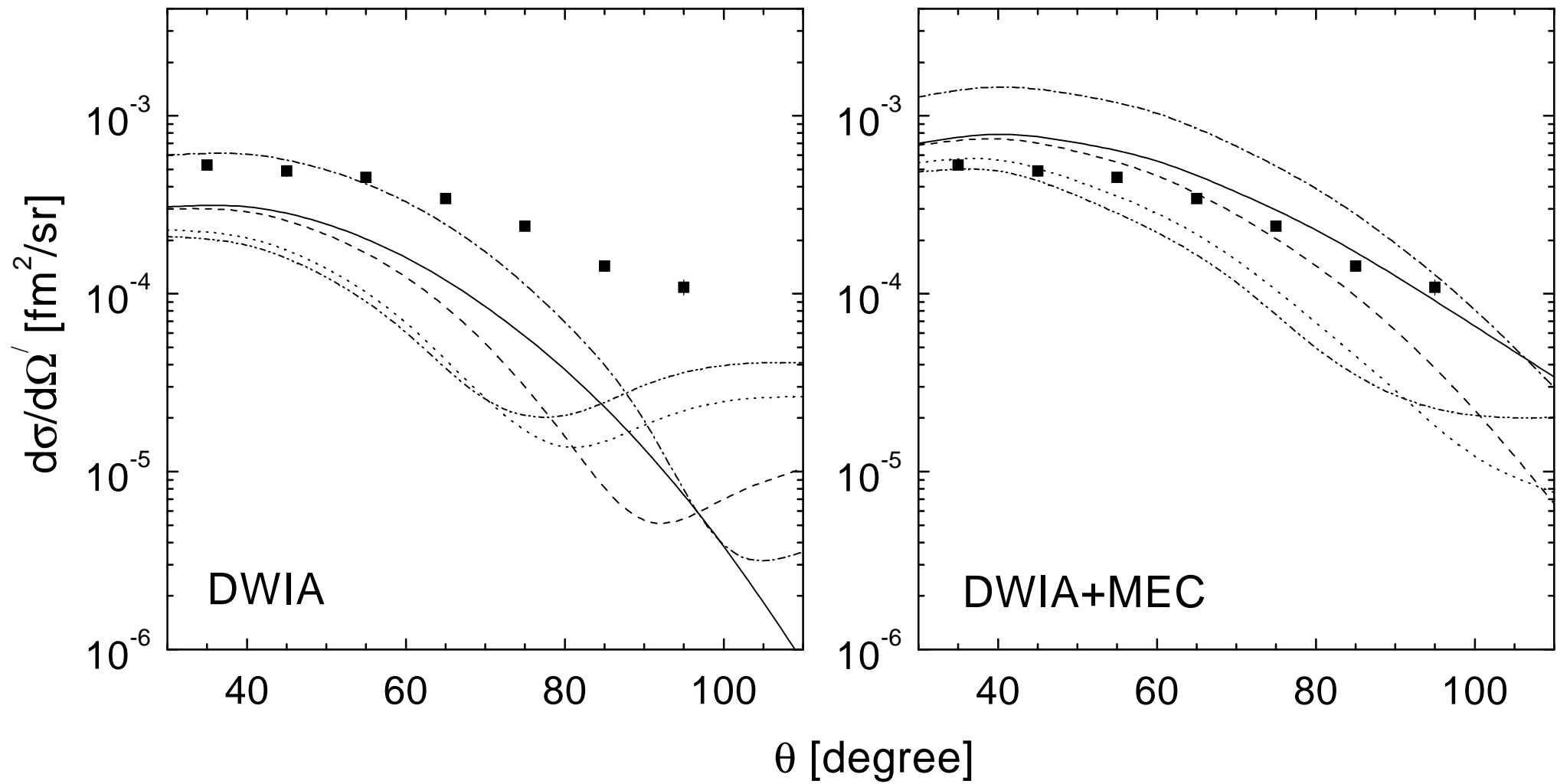
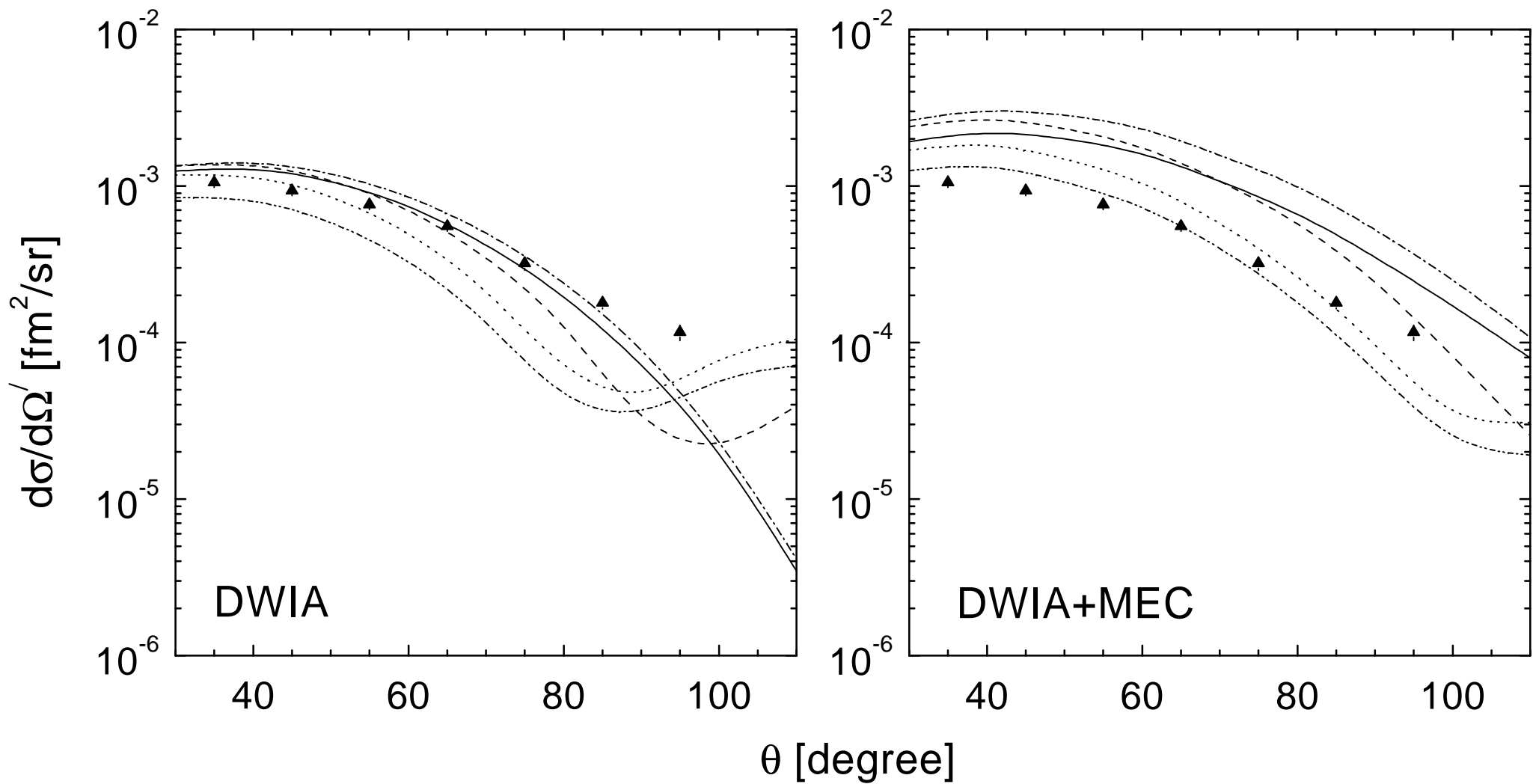
\end{document}